\documentclass[letterpaper,journal]{IEEEtran}

\IEEEoverridecommandlockouts
\usepackage{cite}
\usepackage{amsmath,amssymb,amsfonts}
\usepackage{algorithmic}
\usepackage{graphicx}
\usepackage{textcomp}
\usepackage{xcolor}
\usepackage{psfrag}
\usepackage{tikz}
\usepackage{multicol}
\usepackage{cuted}
\usepackage{epsfig}
\usepackage{color}
\usepackage{geometry}
\usepackage{makecell}

\usetikzlibrary{arrows.meta, positioning, shapes.multipart}
\def\BibTeX{{\rm B\kern-.05em{\sc i\kern-.025em b}\kern-.08em
    T\kern-.1667em\lower.7ex\hbox{E}\kern-.125emX}}

\newcommand{\llrrparen}[1]{
  \left(\mkern-3mu\left(#1\right)\mkern-3mu\right)}
\begin{document}

\title{PAPR of DFT-s-OTFS with Pulse Shaping

\vspace{-0.2 cm}}
\author{\IEEEauthorblockN{Jialiang Zhu, Sanoopkumar P. S., Arman Farhang
}\\ 
\IEEEauthorblockA{Department of Electronic \& Electrical Engineering, Trinity College Dublin, Ireland \\ \{zhuj3, pungayis, arman.farhang\}@tcd.ie}
\vspace{-0.5 cm}
\thanks{This publication has emanated from research conducted with the financial support of Research Ireland under Grant number 18/CRT/6222, 19/FFP/7005(T) and 21/US/3757. For the purpose of Open Access, the author has applied a CC BY public copyright licence to any Author Accepted Manuscript version arising from this submission.
}
}


\maketitle

\begin{abstract}
Orthogonal Time Frequency Space (OTFS) suffers from high peak-to-average power ratio (PAPR) when the number of Doppler bins is large. To address this issue, a discrete Fourier transform spread OTFS (DFT-s-OTFS) scheme
is employed by applying DFT spreading across the Doppler dimension. This paper presents a thorough PAPR analysis of DFT-s-OTFS in the uplink scenario using different pulse shaping filters and resource allocation strategies.
Specifically, we derive a PAPR upper bound of DFT-s-OTFS with interleaved and block Doppler resource allocation schemes. 
Our analysis reveals that DFT-s-OTFS with interleaved allocation yields a lower PAPR than that of block allocation. Furthermore, we show that interleaved allocation produces a periodic time-domain signal composed of repeated quadrature amplitude modulated (QAM) symbols which simplifies the transmitter design. 
Based on our analytical results, the root raised cosine (RRC) pulse generally results in a higher maximum PAPR compared to the rectangular pulse.
Simulation results confirm the validity of  the derived PAPR upper bounds. 
Furthermore, we also demonstrate through BER simulation analysis that the DFT-s-OTFS gives the same performance as OTFS without DFT spreading. 
\end{abstract}


\vspace{-0.4cm}
\section{Introduction}

Orthogonal Time Frequency Space (OTFS) modulation has emerged as a promising waveform for the next-generation wireless communication systems, particularly in high-mobility scenarios such as vehicular networks, satellite communications, and integrated sensing and communication (ISAC) applications\cite{b1}. Unlike orthogonal frequency-division multiplexing (OFDM), OTFS operates in the delay-Doppler domain, allowing it to fully exploit the time and frequency diversity of the channel and remain resilient to the time variations \cite{b6}. 

OTFS can be interpreted as a hybrid carrier waveform that is single carrier along the delay and multi-carrier along the Doppler dimension \cite{b15}. Multi-carrier modulation schemes are known to have high PAPR\cite{b14} and OTFS is not an exception. Similar to OFDM, the PAPR in OTFS increases with the number of Doppler subcarriers \cite{b4}.
A high PAPR significantly reduces the efficiency of RF power amplifiers, often requiring costly linear amplifiers or substantial power back-off to prevent nonlinear distortion.

PAPR of OTFS is analyzed and compared with OFDM in \cite{b4}. 
Several techniques including precoding, compounding and selective mapping have been proposed to reduce the PAPR in OTFS systems\cite{b18,b22,b24}.
Since OTFS can be viewed as a hybrid carrier waveform,
DFT spreading proposed for OFDM systems in \cite{b5} can be applied along the multi-carrier dimension, i.e., Doppler dimension, to reduce PAPR. 
Hence, the authors in \cite{b8} propose a modulation scheme called discrete Fourier transform spread OTFS (DFT-s-OTFS) to reduce the PAPR in multiuser OTFS systems. In \cite{b8}, each user is assigned a group of consecutive delay-Doppler resource blocks (DDRBs) along the Doppler dimension while occupying the full range of delay resources. Prior to mapping the data symbols to the allocated DDRBs, DFT spreading is applied along the Doppler dimension. While DFT-s-OTFS is introduced in \cite{b8}, the study does not consider the influence of pulse shaping or alternative resource allocation schemes on PAPR performance.  In communication systems, pulse shaping is essential for converting digital signals into analog waveforms that can be transmitted over physical channels. The choice of pulse shaping filter can significantly influence the temporal power distribution of the signal, thereby affecting both its peak and average power characteristics \cite{b25}. Consequently, it is essential to investigate the impact of transmit pulse shaping on PAPR. 

This paper focuses on uplink multi-user DFT-s-OTFS systems and investigates the impact of interleaved and block Doppler allocation schemes on PAPR, considering the effect of pulse shaping. 
We derive the PAPR upper bound expressions for both allocation schemes with rectangular and root raised cosine (RRC) transmit pulses. 
Our analysis shows that DFT-s-OTFS with interleaved allocation achieves a significant PAPR reduction compared to block allocation.
An interesting finding of this paper is that interleaved allocation produces a time-domain transmit signal with periodic repetitions of the original quadrature amplitude modulated (QAM) symbols. This eliminates the need for additional processing and simplifies the transmitter architecture.
Our analysis further demonstrates that RRC pulse leads to an increased PAPR for both allocation schemes compared to the rectangular pulse. 
For the RRC pulse, we also investigate the effect of roll-off factor on the PAPR performance.
Simulation results confirm the validity of our analytical derivations. Finally, our bit error rate (BER) analysis shows that all methods achieve comparable performance. 

\vspace{-0.1cm}
\section{System Model}
We consider the uplink of multi-user OTFS system in which $Q$ users simultaneously communicate with a single antenna base station (BS).
 The transmit data symbols are chosen from a $\mathcal{M}$-QAM constellation with zero mean and unit variance. The data symbols are assumed to be independent and identically distributed (i.i.d.) and are placed on a uniform grid in the delay-Doppler domain with delay and Doppler spacings of $\Delta\tau$, and $\Delta\nu$, respectively. 
 Given a delay-Doppler grid with $M$ delay bins and $N$ Doppler bins, the total duration of an OTFS frame in time is $T=MN\Delta\tau$ and the Doppler spacing is defined as $\Delta\nu=\frac{1}{T}=\frac{1}{MN\Delta\tau}$.
 In this paper, we focus on orthogonal Doppler division multiple access (DoDMA) where users are allocated the entire delay bins and non-overlapping subsets of Doppler bins \cite{b8}. Furthermore, the same number of Doppler subcarriers is assigned to each user, i.e., $K=N/Q$.  
 
 Let ${X}_{{q}}[m,k]$ represent the data symbols of user $q$ at the delay bin $m$ and Doppler bin $k$, where $q=0,\ldots, Q-1, m=0,\ldots,M-1$ and $k=0,\ldots, K-1$. 
For DFT-s-OTFS, each user terminal, $q$, spreads the data symbols across its allocated Doppler subcarriers with a $K$-point DFT operation as 
\begin{equation}\label{equ:sym}
    \widetilde{{X}}_{{q}}[m,n'] =  \frac{1}{\sqrt{K}}\sum_{k=0}^{K-1} {X}_{{q}}[m,k] e^{-j 2\pi \frac{k}{K} n'},
\end{equation}
 where $n'=0,\ldots,K-1$. 
 The resulting signal is then mapped to the DDRBs of user $q$, forming the signal ${\widetilde{X}}_{{q}}'[m,n]$, where $n=0,\ldots, N-1$. In this paper, we consider interleaved and block DoDMA schemes. Accordingly, the transmit signal of user $q$, ${\widetilde{X}}_{{q}}'[m,n]$, is non-zero only for $n=Qk+q$ in the interleaved DoDMA and $n=Qq+k$ in the block DoDMA.
 After Doppler carrier allocation, the delay-time domain signal is obtained by applying inverse discrete Fourier transform (IDFT) to ${\widetilde{X}}_{{q}}'[m,n]$ along the Doppler dimension,
 \begin{equation}\label{equ:seq}
    {X}^\text{DT}_{q}[m ,l] = \frac{1}{\sqrt{N}} \sum_{n=0}^{N-1} \widetilde{{X}}_{{q}}'[m,n] e^{j 2\pi \frac{n}{N}l},
\end{equation}
 where $l=0,\ldots,N-1$. 
 Then, the 2D signal ${X}^\text{DT}_{q}[m ,l]$ is converted into a serial stream, i.e.,  ${x}_{q}^{\text{DT}}[m+lM]={X}^\text{DT}_{q}[m,l]$ and a cyclic prefix that is longer than the delay spread of the channel is appended at the beginning of the stream. Finally, the resulting signal is passed through the pulse-shaping filter $g_\text{tx}(t)$, to form
 the transmit signal of user $q$ in baseband, i.e., 
 \begin{equation}\label{equ:seq_pulse}
    x_q(t) = \sum_{i=0}^{MN-1} {x}_q^\text{DT}[i] g_{\text{tx}}(t - i\Delta\tau).
\end{equation}

It is widely known that pulse shaping
can significantly affect PAPR \cite{b25}. Therefore, in the following sections, we analyze the effect of different pulse shaping filters on the PAPR of DFT-s-OTFS both mathematically and through simulations. The PAPR for the signal $x_q(t)$ is defined as
\begin{align} \label{equ:papr}
\text{PAPR} &= \frac{\max\limits_{0\leq t\leq T} \{ |x_q(t)|^2 \}}{\frac{1}{T}  \int_0^T|x_q(t)|^2  dt},
\end{align}
where $\max\limits_{0\leq t\leq T} \{ |x_q(t)|^2 \}$ is the peak power and $\frac{1}{T}  \int_0^T|x_q(t)|^2  dt$ is the average power of $x_q(t)$. Note that the RF modulator increases the PAPR of the baseband signal by 3~dB \cite{b16}. For simplicity and without loss of generality, all derivations and numerical results in this paper consider the baseband signal.  

\section{PAPR Analysis for Interleaved DoDMA}
In this section, we analyze the PAPR of DFT-s-OTFS with interleaved DoDMA. We also derive upper bound PAPR expressions for both rectangular and RRC pulse shaping filters.
\subsection{Rectangular Pulse Shaping}
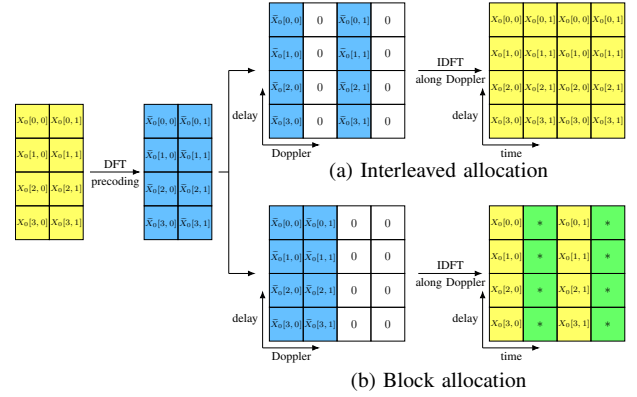
\begin{figure}
\centering
\[
\scalebox{0.45}{
\begin{tikzpicture}[>=Latex, thick]

\definecolor{blue}{RGB}{0, 150, 255}
\definecolor{yellow}{RGB}{255, 255, 0}
\definecolor{green}{RGB}{0, 255, 0}
\foreach \j/\x in {0/0,1/1,2/2,3/3} {
\foreach \i/\x in {0/0,1/1} {
\fill[yellow!60] (1*\i+1.5,-1*\j) rectangle (1*\i+1+1.5,-1*\j+1);
}}
\foreach \j/\x in {0/0,1/1,2/2,3/3} {
\foreach \i/\x in {0/0,1/1} {
  \draw (1*\i+1.5,-1*\j) rectangle (1*\i+1+1.5,-1*\j+1);
  \node at (1*\i+0.5+1.5,0.5-1*\j) {\scriptsize \(X_{0}[\j,\i]\)};
}}

\draw[->] (2.5+1.2,-1) -- (4.0+1.2,-1) node[midway,above ] {DFT}
node[midway,below ] {precoding};

\foreach \j/\x in {0/0,1/1,2/2,3/3} {
\foreach \i/\x in {0/0,1/1} {
  \fill[blue!60] (4.5+1*\i+0.8,0-1*\j) rectangle (4.5+1*\i+1+0.8,0-1*\j+1);
}}

\foreach \j/\x in {0/0,1/1,2/2,3/3} {
\foreach \i/\x in {0/0,1/1} {
  \draw (4.5+1*\i+0.8,0-1*\j) rectangle (4.5+1*\i+1+0.8,-1*\j+1);
  \node at (1*\i+5+0.8,0.5-1*\j) {\scriptsize \(\widetilde{X}_{0}[\j,\i]\)};
}}




\draw[-] (7.5,-1) -- (7.8,-1);
\draw[-] (7.8,-1) -- (7.8,2);
\draw[-] (7.8,-1) -- (7.8,-4);
\draw[->] (7.8,2) -- (8.6,2);
\draw[->] (7.8,-4) -- (8.6,-4) ; 
\foreach \j/\x in {0/0,1/1,2/2,3/3} {
\foreach \i/\x in {0/0,1/1} {
  \fill[blue!60] (9+2*\i,3-1*\j) rectangle (9+2*\i+1,3-1*\j+1);
  \node at (2*\i+9.5,3-1*\j+0.5) {\scriptsize \(\widetilde{X}_{0}[\j,\i]\)};
}}

\foreach \j/\x in {0/0,1/1,2/2,3/3} {
\foreach \i/\x in {0/0,1/1,2/2,3/3} {
  \draw (9+1*\i,3-1*\j) rectangle (9+1*\i+1,3-1*\j+1);
}}

\foreach \j/\x in {0/0,1/1,2/2,3/3} {
\foreach \i/\x in {1/1,3/3} {
;
  \node at (1*\i+9.5,3-1*\j+0.5) {\(0\)};
}}

\foreach \j/\x in {0/0,1/1,2/2,3/3} {
\foreach \i/\x in {0/0,1/1} {
  \fill[blue!60] (9+1*\i,-3-1*\j) rectangle (9+1*\i+1,-3-1*\j+1);
  \node at (1*\i+9.5,-3-1*\j+0.5) {\scriptsize \(\widetilde{X}_{0}[\j,\i]\)};
}}

\foreach \j/\x in {0/0,1/1,2/2,3/3} {
\foreach \i/\x in {0/0,1/1,2/2,3/3} {
  \draw (9+1*\i,-3-1*\j) rectangle (9+1*\i+1,-3-1*\j+1);
}}

\foreach \j/\x in {0/0,1/1,2/2,3/3} {
\foreach \i/\x in {2/2,3/3} {
;
  \node at (1*\i+9.5,-3-1*\j+0.5) {\(0\)};
}}

\draw[->] (13.4,2) -- (15.3,2) node[midway,above ] {IDFT}
node[midway,below ] {along Doppler};
\draw[->] (13.4,-4) -- (15.3,-4) node[midway,above ] {IDFT}
node[midway,below ] {along Doppler}; 

\foreach \j/\x in {0/0,1/1,2/2,3/3} {
\foreach \i/\x in {0/0,1/1,2/2,3/3} {
  \fill[yellow!60] (15.5+1*\i,3-1*\j) rectangle (15.5+1*\i+1,3-1*\j+1);
}}
\foreach \j/\x in {0/0,1/1,2/2,3/3} {
\foreach \i/\x in {0/0,1/1} {
  \node at (1*\i+16,3-1*\j+0.5) {\scriptsize \({X}_{0}[\j,\i]\)};
}}
\foreach \j/\x in {0/0,1/1,2/2,3/3} {
\foreach \i/\x in {0/0,1/1} {

  \node at (1*\i+16+2,3-1*\j+0.5) {\scriptsize \({X}_{0}[\j,\i]\)};
}}

\foreach \j/\x in {0/0,1/1,2/2,3/3} {
\foreach \i/\x in {0/0,1/1,2/2,3/3} {
  \draw (15.5+1*\i,3-1*\j) rectangle (15.5+1*\i+1,3-1*\j+1);
}}


\foreach \j/\x in {0/0,1/1,2/2,3/3} {
\foreach \i/\x in {0/0,1/1} {
  \fill[yellow!60] (15.5+2*\i,-3-1*\j) rectangle (15.5+2*\i+1,-3-1*\j+1);
    \node at (2*\i+16,-3-1*\j+0.5) {\scriptsize \({X}_{0}[\j,\i]\)};
}}

\foreach \j/\x in {0/0,1/1,2/2,3/3} {
\foreach \i/\x in {1/1,3/3} {
  \fill[green!60] (15.5+1*\i,-3-1*\j) rectangle (15.5+1*\i+1,-3-1*\j+1);
  \node at (1*\i+16,-3-1*\j+0.5) {\(*\)};
}}

\foreach \j/\x in {0/0,1/1,2/2,3/3} {
\foreach \i/\x in {0/0,1/1,2/2,3/3} {
  \draw (15.5+1*\i,-3-1*\j) rectangle (15.5+1*\i+1,-3-1*\j+1);
}}

\node at (14,-1) {\LARGE (a) Interleaved allocation};
\node at (14,-7.2) {\LARGE (b) Block allocation};


\draw[->] (8.8,-0.2) -- (10.5,-0.2) node[midway,below ] { Doppler};
\draw[->] (8.8,-0.2) -- (8.8,1.5) node[midway,left ] { delay};

\draw[->] (15.3,-0.2) -- (17,-0.2) node[midway,below ] { time};
\draw[->] (15.3,-0.2) -- (15.3,1.5) node[midway,left ] { delay};

\draw[->] (8.8,-6.2) -- (10.5,-6.2) node[midway,below ] { Doppler};
\draw[->] (8.8,-6.2) -- (8.8,-4.5) node[midway,left ] { delay};

\draw[->] (15.3,-6.2) -- (17,-6.2) node[midway,below ] { time};
\draw[->] (15.3,-6.2) -- (15.3,-4.5) node[midway,left ] { delay};

\end{tikzpicture}}
\]
\vspace{-0.6cm}
\caption{DFT-s-OTFS of different DoDMA schemes for $Q=2, M=N=4$.}
\vspace{-0.6cm}
\label{fig:allocation}
\end{figure}
We firstly consider the rectangular transmit pulse,
 \begin{equation}
g_{\text{tx}}(t) =
\begin{cases} 
1, & 0 \leq t < \Delta\tau \\
0, & \text{otherwise}.
\end{cases}\label{eqn:rect_pulse}
\end{equation}
In interleaved allocation, each user's transmit symbols are spaced $Q$ bins apart along the Doppler dimension, see Fig.~\ref{fig:allocation}(a). Using (\ref{equ:sym}) in (\ref{equ:seq}) and replacing $n$ with $Qk+q$, the transmit signal for DFT-s-OTFS with interleaved allocation, for a given user $q$, can be obtained as
 \begin{align}\label{equ:interleave1}
    {X}_{q}^\text{DT}[m,l] &= \frac{1}{\sqrt{N}} \sum_{k=0}^{K-1} \widetilde{{X}}_{{q}}'[m,Qk+q] e^{j 2\pi \frac{Qk+q}{N} l} \nonumber\\&= \frac{1}{\sqrt{N}} e^{j 2\pi \frac{q}{N}l} \sum_{k=0}^{K-1} \widetilde{{X}}_{{q}}[m,k] e^{j 2\pi \frac{k}{K} l} \nonumber
    \nonumber\\
    &= \frac{1}{\sqrt{N}} e^{j 2\pi \frac{q}{N}l} \frac{1}{\sqrt{K}}\sum_{\kappa=0}^{K-1}{{X}}_{{q}}[m,\kappa] \sum_{k=0}^{K-1} e^{j 2\pi \frac{k}{K} (l-\kappa)}
    \nonumber\\&= \frac{1}{\sqrt{N}} e^{j 2\pi \frac{q}{N}l} \frac{1}{\sqrt{K}}\sum_{\kappa=0}^{K-1}{{X}}_{{q}}[m,\kappa] \left(K\delta[\llrrparen{l}_K-\kappa]\right) 
    \nonumber\\&= \frac{1}{\sqrt{Q}} e^{j 2\pi \frac{q}{N}l} {{X}}_{{q}}[m,\llrrparen{l}_K],
\end{align}
where $\llrrparen{\cdot}_K$ is the modulo $K$ operator and $\delta[\cdot]$ is the discrete Dirac delta function. 
From (\ref{equ:interleave1}), one may realize that the delay-time domain signal ${X}_{q}^\text{DT}[m,l]$ in a given delay bin $m$ is formed by $Q$ repetitions of the input QAM symbols ${X}_{{q}}[m,k]$ for $k=0,\ldots,K-1$ that are scaled by $1/\sqrt{Q}$ and modulated to the Doppler frequency $q/N$. The modulation stage can be absorbed into RF modulator.
Consequently, interleaved DoDMA scheme substantially reduces the transmitter complexity, making it highly attractive for practical implementation. 

To derive the upper bound for PAPR using (\ref{equ:papr}), the maximum possible peak power and the average power for $x_q(t)$ must be calculated. 
To this end, the maximum peak power of the transmit signal can be obtained as 
\begin{align}\label{interleave_max}
\max_{0\leq t\leq T} \left\{ |x_q(t)|^2 \right\} &= \max_{m,l} \left\{ \left|X_q^{\text{DT}}[m,l] \right|^2 \right\}\nonumber\\&
=\max_{m,l} \left\{ \left| \frac{1}{\sqrt{Q}} {X}_{{q}}[m,\llrrparen{l}_K] \right|^2 \right\},
\end{align}
where ${X}_{{q}}[m,\llrrparen{l}_K]$ takes values from a $\mathcal{M}$-QAM constellation with the maximum power $\frac{3(\sqrt{\mathcal{M}}-1)^2 }{(\mathcal{M}-1)}$, \cite{b11}. Hence,  
\begin{equation}\label{interleave_max1}
\max_{0\leq t\leq T} \left\{ |x_q(t)|^2 \right\} \leq \frac{3(\sqrt{\mathcal{M}}-1)^2 }{Q(\mathcal{M}-1)}.
\end{equation}
The average power of the transmit signal, $x_q(t)$ can be calculated as 
\begin{align}\label{interleave_p_avg}
P_{\text{avg}} &= \frac{1}{MN}  \sum_{i=0}^{MN-1} \mathbb{E} \left\{ \left| {x}^{\text{DT}}_q[i] \right| ^2 \right\} 
\nonumber\\&= \frac{1}{MN}  \sum_{m=0}^{M-1}Q\sum_{k=0}^{K-1} \mathbb{E} \left\{ \left| \frac{1}{\sqrt{Q}} {X}_{{q}}[m,k] \right| ^2 \right\} 
= \frac{1}{{Q}},
\end{align}
where $\mathbb{E} \left\{ \left|  {X}_{{q}}[m,k] \right| ^2 \right\}=1$. Using (\ref{interleave_max1}) and (\ref{interleave_p_avg}), the upper bound for PAPR of the DFT-s-OTFS transmit signal with interleaved DoDMA at each user terminal is obtained as 
\begin{equation}\label{papr_dft1}
\text{PAPR}_\text{interleaved}^\text{DFT-s-OTFS}\leq \frac{3(\sqrt{\mathcal{M}}-1)^2 }{(\mathcal{M}-1)}.
\end{equation}
It can be observed that the upper bound PAPR of interleaved DoDMA with rectangular pulse is determined by the QAM constellation. However, rectangular pulse shaping is not practically feasible as it requires an infinite bandwidth \cite{book1}.

\vspace{-0.2cm}
\subsection{RRC Pulse Shaping} 
Raised cosine (RC) pulse shaping is utilized in modern communication systems due to its favorable spectral behavior and reduced intersymbol interference. An RRC filter is employed at the transmitter and a matched filter is utilized at the receiver to achieve an effective RC filter response between the transmitter and the receiver \cite{book1}. In this section, we analyze the PAPR of the interleaved DFT-s-OTFS with an RRC transmit filter. The time-domain representation of RRC pulse is given by
\begin{equation}\label{eqn:rrc}
    g_{\text{tx}}(t) =  \frac{4\beta}{\pi }\frac{ \left[ \cos\left( (1 + \beta) \pi t / \Delta\tau \right) + \frac{\sin\left( (1 - \beta) \pi t / \Delta\tau \right)}{4\beta t / \Delta\tau} \right]}{1 - \left( 4\beta t / \Delta\tau \right)^2},
\end{equation}
where the parameter $0\leq\beta\leq1$, known as the roll-off factor, governs both the bandwidth and the temporal shape of the pulse. Its frequency response can be represented as\cite{b26}
\begin{equation}
G(f) \!=\!
\begin{cases}
{\Delta\tau}, & 0  \! \leq \!  |f|  \! \leq \!  \frac{1 - \beta}{2\Delta\tau} \\[7pt]
\! \frac{\Delta\tau}{\sqrt{2}}\sqrt{ \left[\! 1 \!+\! \cos \!\left( \!\frac{\pi \Delta\tau}{\beta} \! \left( \! |f| \!-\! \frac{1 - \beta}{2\Delta\tau} \! \right) \! \right)\! \right] }, & \frac{1 - \beta}{2\Delta\tau} \! \leq \! |f| \! \leq \! \frac{1 + \beta}{2\Delta\tau} \\[7pt]
0, & |f| \! > \! \frac{1 + \beta}{2\Delta\tau}
\end{cases}
\end{equation}
From (\ref{equ:interleave1}), the serial stream $x_q^{\text{DT}}[i]$ in (\ref{equ:seq_pulse}) includes the QAM transmit symbols modulated to each user's Doppler frequency, i.e., $q/N$. The input signal to the pulse-shaping filter $g_{\text{tx}}(t)$ can be modeled as an impulse train with non-zero values at the delay spacings $\Delta\tau$ with the amplitude $x_q^{\text{DT}}[i]$, i.e., $s(t)=\sum_{i=0}^{MN-1}x_q^\text{DT}[i]\delta(t-i\Delta\tau)$. Thus, (\ref{equ:seq_pulse}) can be rearranged as
 \begin{equation}\label{eqn:complex}
    x_q(t) = s(t) * g_{\text{tx}}(t ).
\end{equation}
Employing Parseval's theorem and using the duality of convolution and multiplication in time and frequency domains, respectively, the average power of transmit signal can be calculated as
\begin{equation}\label{equ:p_avg_rrc}
P_{\text{avg}} = \mathbb{E} \left\{ |{x_q}(t)|^2 \right\} = \int_{-\infty}^{\infty} \phi_{ss}(f) |G(f)|^2 \, df,
\end{equation}
where $\phi_{ss}(f)$ is the power spectral density function of $s(t)$. 
As we consider uncorrelated QAM symbols with zero mean and unit power, using (\ref{equ:interleave1}),
$\mathbb{E} \left[ |{x}_{q}^{\text{DT}}[i]|^2 \right] =1/Q$. From Parseval's theorem and periodicity of $\delta(t-i\Delta\tau)$ in $s(t)$,
$\phi_{ss}(f)=\frac{1}{\Delta\tau}{\mathbb{E} \left\{ |{x}_{q}^{\text{DT}}[i]|^2 \right\}}=\frac{1}{Q\Delta\tau}$. 
Since, the average power of the RRC filter is $\Delta \tau$ \cite{b26},
the average power for interleaved DFT-s-OTFS with RRC pulse shaping is obtained as
\begin{equation}
P_{\text{avg}}=\frac{1}{Q\Delta\tau}\int_{-\infty}^{\infty} |G(f)|^2 \, df =\frac{1}{Q}.
\end{equation}
Given the inequality ${x}_q^\text{DT}[i] g_{\text{tx}}(t - i\Delta\tau) \!\leq\! \left|{x}_q^\text{DT}[i]\right|\cdot \left| g_{\text{tx}}(t \!-\! i\Delta\tau)\right|$ and assuming all the transmit symbols have maximum power, the upper bound of peak power can be obtained by
\begin{align}{\label{peak_pulse_power}}
\max_t \left\{ |{x}_q[t]|^2 \right\} &= \max_t \left\{ \left|\sum_{i=0}^{MN-1} {x}_q^\text{DT}[i] g_{\text{tx}}(t - i\Delta\tau)\right|^2 \right\}
\nonumber \\ &\leq  \max_t \left\{ \left|\sum_{i=0}^{MN-1} \left|{x}_q^\text{DT}[i]\right| \cdot \left|g_{\text{tx}}(t - i\Delta\tau)\right| \right|^2 \right\}
\nonumber \\ &\leq  g_0^2\max_i \left|{x}_q^\text{DT}[i] \right|^2 ,
\end{align}
where $g_0=\max_t \left\{\sum_{i = 0}^{MN-1}|g_{\text{tx}}(t - i\Delta\tau)|\right\} $. However, the value of $g_0$ locates at different time indices depending on the roll-off factor. Specifically, from \cite{b26}, if $0 \leq \beta \leq 0.4$, the maximum value is obtain at $t=\Delta\tau/2$ and
\begin{equation}\label{equ:rrc_max1}
g_0\!= \sum_{i=0}^{L_\text{span}-1}\frac{A(i) + B(i)}{-\frac{\pi}{2}(1 + 2i)\left(1 - \left(2(1 + 2i)\beta\right)^2\right)}, 
\end{equation}
where $L_\text{span}$ is the span of upsampled RRC,  $A(i) = (-1)^{i+1} \cos\left( \beta \frac{\pi}{2} (1 + 2i) \right)$, $B(i) = 2\beta(1 + 2i)(-1)^{i} \sin\left( \beta \frac{\pi}{2} (1 + 2i) \right)$.
If $\beta >0.4$, the maximum value is obtained at $t=0$ and
\begin{equation}\label{equ:rrc_max2}
g_0\!= 1 - \beta + \frac{4\beta}{\pi} + \sum_{i=0}^{L_\text{span}-1} \left| \frac{C(i) + D(i)}{\frac{\pi}{2}(2i+1) \left(1 - (2(2i+1)\beta)^2 \right)} \right| ,
\end{equation}
where $C(i) = (-1)^{i} \sin(\beta i \pi)$, $D(i) = 4(i+1)\beta (-1)^{i+1} \cos(\beta (i+1) \pi)$.
Therefore, according to (\ref{peak_pulse_power}), the maximum peak power is bound by $\frac{3(\sqrt{\mathcal{M}}-1)^2}{Q(\mathcal{M}-1)}g_0^2$.
Thus, the upper bound of PAPR of DFT-s-OTFS of interleaved DoDMA considering RRC pulse can be derived as 
\begin{equation}{\label{papr_inter_pulse}}
\text{PAPR}_\text{interleaved-RRC}^\text{DFT-s-OTFS}\leq \frac{3g_0^2(\sqrt{\mathcal{M}}-1)^2}{(\mathcal{M}-1)}
\end{equation}
Comparing (\ref{papr_inter_pulse}) with (\ref{papr_dft1}), it can be observed that the RRC pulse leads to an increased PAPR upper bound by a factor of $g_0^2$ which depends on the roll-off factor. 

\vspace{-0.3cm}
\section{PAPR Analysis for Block DoDMA}\label{block}
In this section, we derive PAPR upper bound expressions for DFT-s-OTFS with block DoDMA using both rectangular and RRC transmit pulses.
\vspace{-0.6cm}
\subsection{Rectangular Pulse Shaping}
For block DoDMA scheme, as illustrated in Fig.~\ref{fig:allocation}(b), the transmit symbols of each user occupy $K$ consecutive Doppler bins and all the delay bins. 
Substituting (\ref{equ:sym}) into (\ref{equ:seq}) and replacing $n$ with $Qq+k$, the transmit signal of a given user $q$ for DFT-s-OTFS with block allocation can be written as
{
 \begin{align}\label{block1}
    {X}_q^\text{DT}[m,l] &=\! \frac{1}{\sqrt{N}} \sum_{k=0}^{K-1} \widetilde{{X}}_{{q}}'[m,Qq+k] e^{j 2\pi \frac{Qq+k}{N} l}
   \nonumber\\& = \! \frac{1}{\sqrt{N}} e^{j 2\pi \frac{q}{K}l} \sum_{k=0}^{K-1} \widetilde{{X}}_{{q}}[m,k] e^{j 2\pi \frac{k}{N} l}
    \nonumber\\& = \! \frac{1}{\sqrt{N}} e^{j 2\pi \frac{q}{K}l} \sum_{\kappa=0}^{K-1} \!\frac{1}{\sqrt{K}}{{X}}_{{q}}[m,\kappa]\sum_{k=0}^{K-1}  e^{j 2\pi \frac{k}{N} (l-Q\kappa)}.
\end{align}
}
If $l=Q\kappa$, the transmit signal will be $\frac{1}{\sqrt{Q}}{X}_{{q}}[m,\kappa]e^{j 2\pi \frac{Qq}{K}\kappa}$,
where $\kappa=0,\ldots,K-1$.
Thus, the time domain signal with block allocation at the time indices $l=Q\kappa$ for a given user, $q$, is the original QAM symbols scaled by $1/\sqrt{Q}$ and phase shifted by $e^{j 2\pi \frac{Qq}{K}\kappa}$. For the indices $l\neq Q\kappa$, the signal is the superposition of all input QAM symbols within the block, each subject to different complex weightings as shown in Fig.~\ref{fig:allocation}(b). 

Using (\ref{block1}) and $\mathbb{E} \left\{ \left|  {X}_{{q}}[m,\kappa] \right| ^2 \right\}=1$, the average power of the transmit signal can be obtained as
\begin{align}\label{block_p_avg}
P_{\text{avg}} &= \frac{1}{MN}  \sum_{m=0}^{M-1}\sum_{l=0}^{N-1} \mathbb{E} \left\{ \left| {X}^{\text{DT}}_{{q}}[m,l] \right| ^2 \right\} 
\nonumber \\&=\frac{1}{{NK}}K^2  = \frac{1}{{Q}}.
\end{align}

Using (\ref{block1}), and Cauchy-Schwarz inequality, the maximum peak power of the transmit signal is bounded by 
{\small
\begin{align}\label{block_max}
\max_{0\leq t\leq T} \left\{ |x_q(t)|^2 \right\} &= \max_{m,l} \left\{ \left|X_q^{\text{DT}}[m,l] \right|^2 \right\}
\nonumber\\&\leq\!\! \frac{1}{{NK}}\!\!\left( \sum_{\kappa=0}^{K-1}\!\! \left| X_q[m,\kappa] \right|^2 \!\!\right)\!\!\! \left( \sum_{\kappa,k=0}^{K-1} \!\!\left| e^{-j2\pi  \frac{k}{K}(l-Q\kappa)} \right|^2 \!\!\right)
\nonumber\\&=\frac{1}{N}K^2\frac{3(\sqrt{\mathcal{M}}-1)^2 }{(\mathcal{M}-1)}=\frac{3K(\sqrt{\mathcal{M}}-1)^2 }{Q(\mathcal{M}-1)}.
\end{align}
}
Therefore, the PAPR upper bound of the DFT-s-OTFS with block DoDMA at each user is obtained as
\begin{equation}\label{papr_block}
\text{PAPR}_\text{block}^\text{DFT-s-OTFS}\leq \frac{3K(\sqrt{\mathcal{M}}-1)^2 }{(\mathcal{M}-1)}.
\end{equation}
This upper bound matches the derivation in \cite{b8}. If we compare (\ref{papr_block}) with (\ref{papr_dft1}), DFT-s-OTFS with block allocation has a PAPR of $K$ times compared with interleaved allocation.  

\vspace{-0.5cm}
\subsection{RRC Pulse Shaping}
For block DoDMA with RRC pulse, the transmit signal can also be represented using (\ref{eqn:complex}). As proven in (\ref{block_p_avg}), $\mathbb{E} \left[ |{x}_{q}^{\text{DT}}[i]|^2 \right] \!\!=\!\!1/Q$. Thus,
$\phi_{ss}(f)\!=\!\frac{1}{\Delta\tau}{\mathbb{E} \left\{ |{x}_{q}^{\text{DT}}[i]|^2 \right\}}\!=\!\frac{1}{Q\Delta\tau}$. 
According to (\ref{equ:p_avg_rrc}), the average power of transmit signal is 
\begin{equation}\label{equ:p_avg_rrc2}
P_{\text{avg}} = \mathbb{E} \left\{ |{x_q}(t)|^2 \right\} = \frac{1}{Q}.
\end{equation}
The peak power of DFT-s-OTFS with RRC pulse for block allocation is also bounded by (\ref{peak_pulse_power}).
Using (\ref{equ:rrc_max1}), (\ref{equ:rrc_max2}) and (\ref{block_max}), the maximum peak power is $\frac{3g_0^2K(\sqrt{\mathcal{M}}-1)^2}{Q(\mathcal{M}-1)}$. Thus, the upper bound on PAPR for DFT-s-OTFS for block DoDMA considering RRC pulse shaping is obtained as
\begin{equation}{\label{papr_block_pulse}}
\text{PAPR}_\text{block-RRC}^\text{DFT-s-OTFS}\leq\frac{3g_0^2K(\sqrt{\mathcal{M}}-1)^2}{(\mathcal{M}-1)}.
\end{equation}
Comparing (\ref{papr_block_pulse}) with (\ref{papr_inter_pulse}), we can observe that with block allocation, the PAPR of the DFT-S-OTFS increases by a factor of $K$ compared to interleaved DoDMA. 

\section{Simulation Results}
In this section, we numerically evaluate the PAPR performance of DFT-s-OTFS with pulse-shaping for block and interleaved DoDMA while validating our analytical results. 
If not mentioned otherwise, we consider $M=128$, $N=32$, $Q=4$, $K=8$ and 16 QAM to modulate the information bits. For the bit error rate (BER) performance analysis, the Extended Vehicular A (EVA) channel model \cite{b9} at the velocity of $v\!=\!500$~km/h is used. Perfect channel state information (CSI) is assumed at the receiver, and the minimum mean square error (MMSE) equalizer is employed for detection.
\begin{figure}
  \centering
  \begin{minipage}[t]{0.235\textwidth}
    \includegraphics[width=\textwidth]{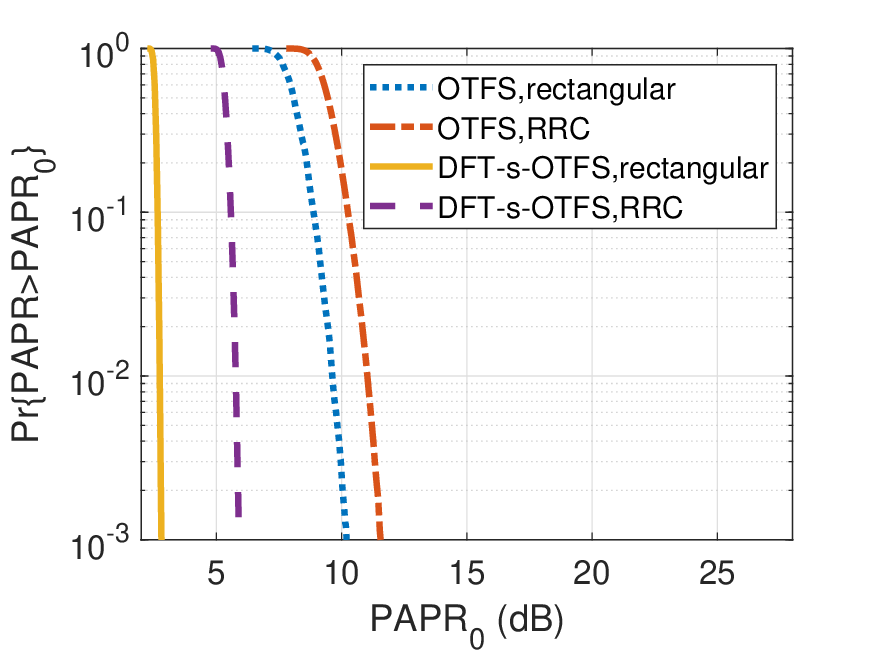}
    \centering{\footnotesize  {(a) Interleaved allocation}}
    \label{fig:sub1}
  \end{minipage}
  \begin{minipage}[t]{0.235\textwidth}
    \includegraphics[width=\textwidth]{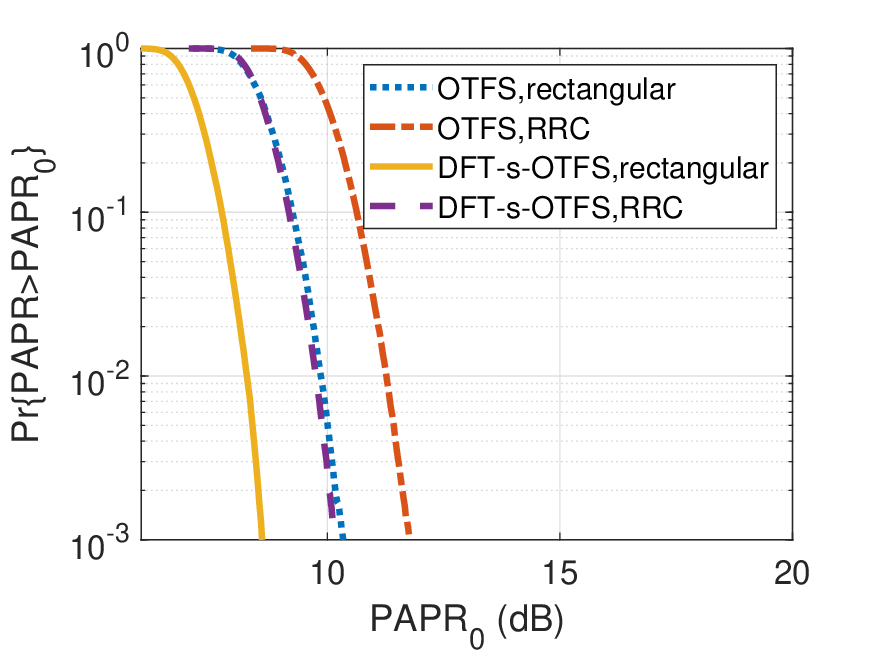}
    \centering{\footnotesize {(b) Block allocation}}
    \label{fig:sub2}
  \end{minipage}
  \caption{CCDF of PAPR of DFT-spread OTFS and OTFS for different allocation schemes, with rectangular and root raised cosine pulse shaping of $\beta=0.5$.}
  \label{fig:compare}
\end{figure}

Fig.~\ref{fig:compare} presents the complementary cumulative distribution function (CCDF) of the PAPR for DFT-s-OTFS and conventional OTFS with interleaved and block DoDMA for rectangular and RRC pulse shaping. 
It can be observed from Fig.~\ref{fig:compare}(a), that the PAPR of DFT-s-OTFS with rectangular pulse-shaping and interleaved DoDMA consistently falls below 2.55 dB, which is the upper bound calculated in (\ref{papr_dft1}). Compared to OTFS with the interleaved allocation, DFT-s-OTFS exhibits 8 dB PAPR improvement. Using the RRC pulse with a roll-off factor of 0.5, the upper bound increases to 6.03~dB according to  (\ref{papr_inter_pulse}), which is also observed in Fig.~\ref{fig:compare}(a). 

For block allocation with rectangular pulse shaping, the upper bound on PAPR using  (\ref{papr_block}) is 11.25 dB. It can be seen from  Fig.~\ref{fig:compare}(b) that simulation results remain below the derived upper bound.  Furthermore, it can also be observed from the figure that compared to OTFS without DFT spreading, for block allocation, the DFT-s-OTFS has a 2 dB gain. With RRC pulse shaping, the upper bound increases to 15.06 dB from (\ref{papr_block_pulse}). The rise in PAPR for both schemes under RRC is due to overlapping pulse contributions in time, where aligned phases can cause significant instantaneous power spikes. Notably, with rectangular pulse, interleaved DoDMA achieves 6 dB lower PAPR than block DoDMA.

\begin{figure}
  \centering
  \begin{minipage}[t]{0.235\textwidth}
    \includegraphics[width=\textwidth]{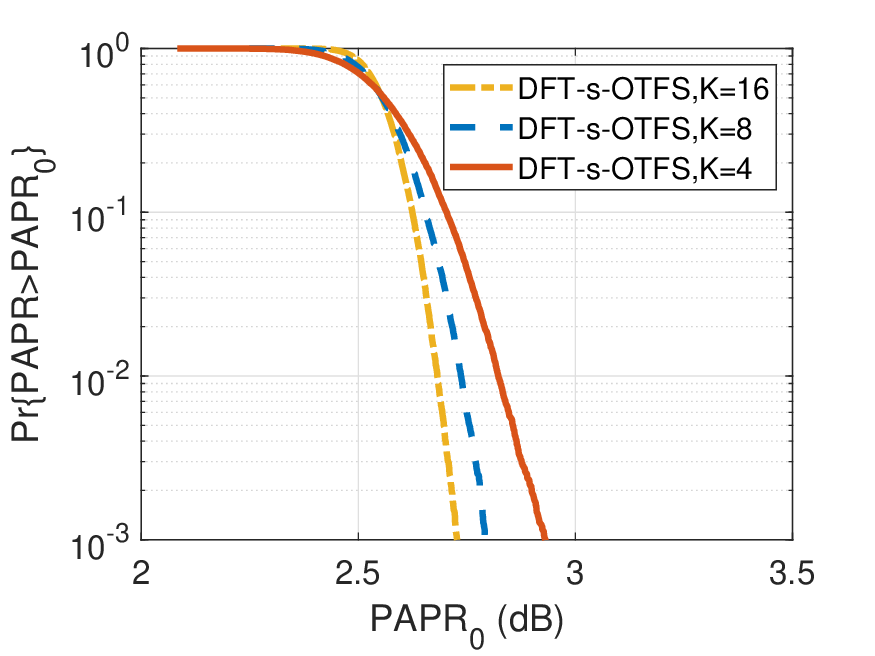}
    \centering{\footnotesize  {(a) Interleaved allocation}}
    \label{fig:sub11}
  \end{minipage}
  \begin{minipage}[t]{0.235\textwidth}
    \includegraphics[width=\textwidth]{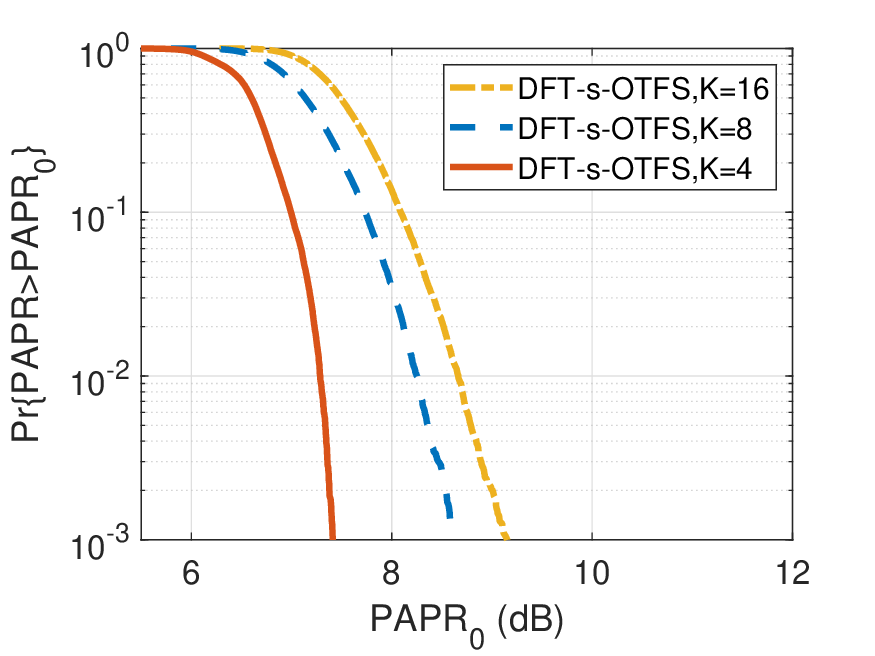}
    \centering{\footnotesize {(b) Block allocation}}
    \label{fig:sub22}
  \end{minipage}
  \caption{CCDF of PAPR of DFT-spread OTFS with different allocation schemes of rectangular transmit pulse using different $K$.}
  \vspace{-0.2cm}
  \label{fig:compare_k_n}
\end{figure}

In Fig.~\ref{fig:compare_k_n}, we compare the PAPR performance of DFT-s-OTFS under different values of $K$ for a fixed value of $N$. 
From Fig.~\ref{fig:compare_k_n}(a), varying $K$ has a minimal impact of around 0.5 dB on the PAPR of the interleaved DoDMA when $N$ is fixed. In contrast, Fig.~\ref{fig:compare_k_n}(b) demonstrates that increasing $K$ leads to a higher PAPR in block DoDMA. These observations are consistent with the analytical results in (\ref{papr_dft1}) and (\ref{papr_block}), which indicate that the PAPR of DFT-s-OTFS with interleaved DoDMA is almost independent of $K$ while the PAPR of block DoDMA increases linearly with $K$.

\begin{figure}
  \centering
  \begin{minipage}[t]{0.235\textwidth}
    \centering
    \includegraphics[width=\textwidth]{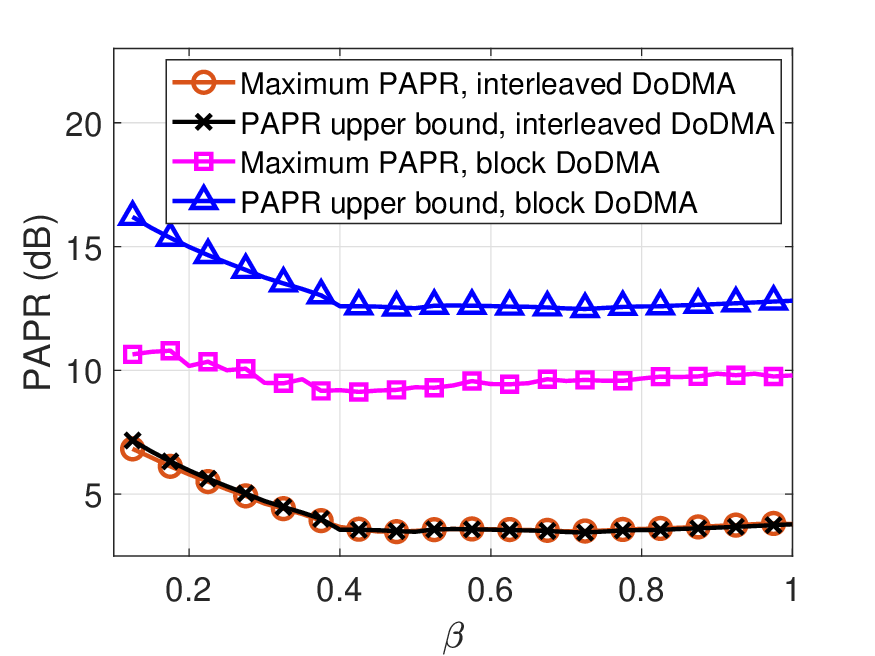}
    \vspace{-0.6cm}
    \caption{PAPR of DFT-s-OTFS versus RRC roll-off factor.}
    \label{fig:ps}
    \vspace{-0.6cm}
  \end{minipage}
  \hfill
  \begin{minipage}[t]{0.235\textwidth}
    \centering
    \includegraphics[width=\textwidth]{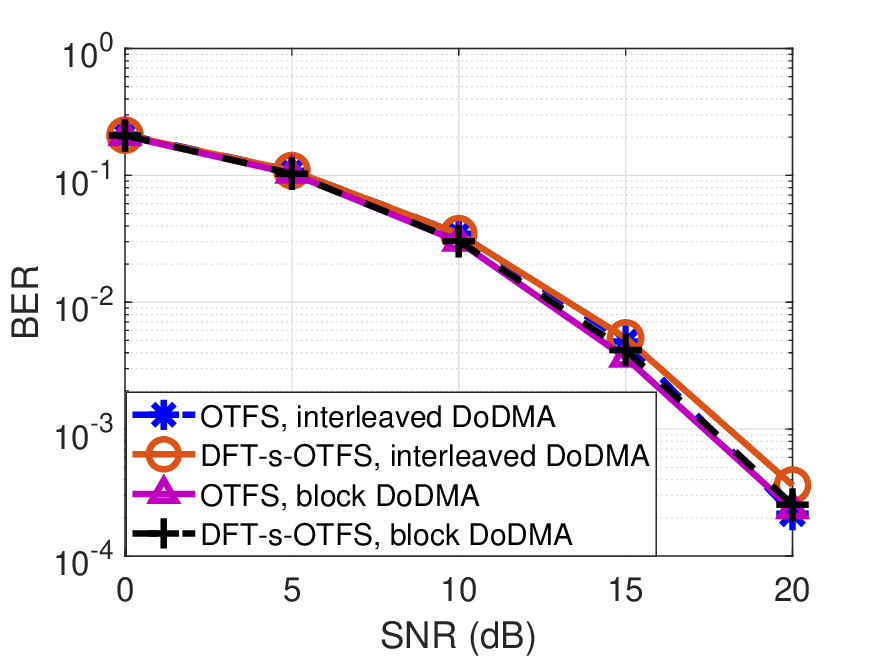}
    \vspace{-0.6cm}
    \caption{BER performance of DFT-s-OTFS and OTFS versus SNR.}
    \label{fig:ber}
  \end{minipage}
\end{figure}

Fig.~\ref{fig:ps} shows a comparison between the analytical upper bound and the simulated maximum PAPR for DFT-s-OTFS under different RRC roll-off factors for $4$-QAM. The upper bounds are computed using (\ref{papr_inter_pulse}) and (\ref{papr_block_pulse}).  
The results show that PAPR decreases as the roll-off factor increases to approximately $\beta=0.4$, and then remains relatively constant, with a slight increase between 0.4 and 1 for both interleaved and block DoDMA. Notably, the upper bound for interleaved DoDMA closely follows the simulated maximum PAPR, indicating its tightness. In contrast, while the simulated PAPR of block DoDMA also remains below the bound, the upper bound is relatively loose.

Fig.~\ref{fig:ber} illustrates the BER performance of DFT-s-OTFS and OTFS under different DoDMA schemes. We used an RRC filter with the roll-off factor of $0.5$ at the transmitter and a matched filter at the receiver. 
The results show that DFT-s-OTFS and OTFS with the interleaved and block DoDMA achieve comparable BER performance. This demonstrates the fact that DFT-s-OTFS with interleaved DoDMA can significantly reduce PAPR with a negligible BER performance loss compared to OTFS or DFT-s-OTFS with block DoDMA.
\vspace{-0.3cm}
\section{Conclusion}
\vspace{-0.1cm}
\label{sec:conclusion}
In this paper, we investigated the impact of resource allocation and pulse shaping on the PAPR in the uplink of DFT-s-OTFS. We showed that DFT-s-OTFS has significantly lower PAPR than OTFS while providing about the same BER performance.  We derived upper bounds on the PAPR of interleaved and block DoDMA schemes with rectangular and RRC pulse shaping which were validated by simulations. The results demonstrate that interleaved DoDMA achieves a substantial reduction in PAPR compared to block DoDMA. Furthermore, the time-domain signal of interleaved DoDMA was found to be repetitions of the original QAM symbols, simplifying the transmitter. We showed that increasing the DFT spreading size has a negligible impact on PAPR in interleaved DoDMA, whereas it increases the PAPR for block DoDMA. We also investigated the effect of the RRC pulse rolloff factor on PAPR
for both block and interleaved DoDMA. Our study
revealed that pulse shaping with the RRC pulse increases PAPR compared to the rectangular pulse which can be mitigated by increasing the roll-off factor.

\bibliographystyle{IEEEtran} 

\end{document}